\documentclass[aps,twocolumn,floats,prd,nofootinbib,superscriptaddress,showkeys,10pt]{revtex4-2}

\usepackage{graphicx} 
\usepackage{amsfonts,amsmath,amssymb} 
\usepackage{hyperref}
\usepackage{enumitem}
\usepackage[english]{babel}
\usepackage{bbold,dsfont}
\usepackage{color}
\newcommand{\vi}[1]{\textcolor{red}{\textbf{V.I.: #1}}}
\newcommand{\os}[1]{\textcolor{blue}{\textbf{O.S.: #1}}}

\newcommand{\eq}[1]{\begin{align} #1 \end{align}}
\newcommand{\mt}{\mathcal{T}}
\begin{document}

\title{Static spherically symmetric configurations with nonlinear scalar field sigma models}
\title{Global and asymptotic properties of self-gravitating nonlinear sigma models}

\author{Oleksandr S.~Stashko}
\email{alexander.stashko@gmail.com}
\affiliation{Black Hole Initiative,
Harvard University, Cambridge, MA, 02138, USA
\looseness=-1}
\affiliation{Institute of Cosmology, Department of Physics and Astronomy, Tufts University, Medford, MA 02155, USA
\looseness=-1}
\author{Valery I.~\surname{Zhdanov}}
\email{valeryzhdanov@gmail.com}
\affiliation{Astronomical Observatory, Taras Shevchenko National University of Kyiv,  Ukraine}
\affiliation{Applied Physics Department, Igor Sikorsky Kyiv Polytechnic Institute, Ukraine}
\author{Mark P.~Hertzberg}
\email{mark.hertzberg@tufts.edu}
\affiliation{Institute of Cosmology, Department of Physics and Astronomy, Tufts University, Medford, MA 02155, USA
\looseness=-1}


\begin{abstract}
We present an analytical study of the global and asymptotic properties of static, spherically symmetric, asymptotically flat solutions in general relativity and its \(D\)-dimensional generalization with multiple minimally coupled scalar fields described by a general nonlinear sigma model. Assuming a complete Riemannian target space and a non-negative scalar potential satisfying suitable growth conditions, we prove that solutions remain regular at every finite nonzero radius and can be extended inward to a naked singularity at the center. In particular, this excludes singularities on spherical shells at finite radius. We further show that all solutions exhibit universal asymptotic behavior near the central singularity. Finally, we demonstrate that these spacetimes can possess at most one photon sphere, which, if present, is necessarily unstable.
 
\end{abstract}


\maketitle%


\section{Introduction} \label{Introduction}
Scalar fields provide a simple extension of General Relativity (GR) and have been widely used in astrophysics, particularly in modeling compact objects \cite{DamourEspositoFarese1993,LieblingPalenzuela2023,Doneva2024Scalarization}. While many applications involve a single canonical scalar field (SF), the inclusion of multiple scalar fields allows for more general interactions, including noncanonical kinetic couplings. Such interactions naturally lead to nonlinear sigma models, in which the scalar fields are described in terms of a  nontrivial target-space metric. The applications of the corresponding models have  attracted considerable interest in recent years \cite{Brax:2023now,Machet:2025vzt,Pereira:2026qpr,Doneva_2020,PhysRevD.102.084032,PhysRevD.104.064045}.

In this paper, we consider static, spherically symmetric (SSS) solutions arising in General Relativity and its $D$-dimensional generalizations with multiple scalar fields. It is natural to assume that sufficiently far from an isolated system (in the weak-field region), we have regular solutions of the Einstein-scalar field equations. Problems may arise when transitioning into the strong-field region, where nonlinear effects can lead to spacetime singularities. The occurrence of singularities in GR is a common phenomenon \cite{Hawking-Ellis1975}. Related problems have been considered in connection with various no-hair theorems \cite{Bekenstein1972,Bekenstein1972a,Doneva_2020,Doneva2025IJMPD,Heusler:1992ss,Herdeiro-Radu2015IJMPD}, as well as cosmic censorship hypotheses \cite{Penrose1965,Penrose:1969pc,Penrose2002}  concerning the possible existence of naked singularities; see also \cite{Wald1980,Who_afraids,Ishibashi-Wald2003}. Counterexamples to cosmic censorship have also been discussed in the literature \cite{Christodoulou1984,Ori1987,Joshi1993,Ong2020}.

In the case of asymptotically flat SSS configurations, one question concerns the possibility of extending solutions from spatial infinity to the entire range of the radial coordinate\footnote{Throughout this paper we use Schwarzschild-like (curvature) coordinates, corresponding to metric (\ref{SS-metric}) given below.} $r>0$, all the way to the center. A well-known example is the Fisher–Janis–Newman–Winicour solution \cite{Fisher,JNW, PhysRevD.40.2564}, see also \cite{Wyman,Agnese1985,Virbhadra_1997}, which describes a SSS configuration with a canonical massless scalar field. This solution is regular for all nonzero values of $r$ but possesses a naked singularity at the center. However, in more general configurations, singularities may occur at finite nonzero radii, preventing the continuation of solutions toward the center. Such spherical-like singularities have been found in various gravitational theories \cite{PhysRevD.67.064009,PhysRevLett.109.251102,Nagy:2021bha,PhysRevD.76.024003,Verbin:2020fzk,Strongly2021}. In particular, SSS solutions with nonlinear scalar potentials may develop spherical singularities already in flat spacetime \cite{ZhdSt}, as well as in GR \cite{Strongly2021}.

It was proved in Ref. \cite{ZhdSt} that SSS solutions with several nonlinear scalar fields with canonical kinetic terms and exponentially bounded monotonic potentials can be extended to all $r>0$. Under these assumptions, gravity suppresses the formation of singularities at finite radii, while a naked singularity may remain at the center. 

In the present paper, we extend the results of Ref. \cite{ZhdSt} in two directions. First, we consider a noncanonical kinetic term corresponding to a nonlinear sigma model with a generally nontrivial target-space metric. Second, we remove the monotonicity requirement on the scalar potential, allowing, in particular, potentials with multiple minima. We establish sufficient conditions for the continuation of asymptotically flat SSS solutions over the entire interval $r>0$ and investigate their asymptotic behavior near the center.

In Section \ref{notations}, we formulate the model and the assumptions on the target-space geometry and scalar potential. In Section \ref{sigma_mod}, we derive the Einstein-scalar field equations for SSS configurations and establish the main results on the continuation of solutions over the interval $(0,\infty)$. Section \ref{Section: asymptotic relations} is devoted to the asymptotic behavior as $r\to 0$. In Section \ref{Discussion}, we discuss  the results and outline their possible applications. In Appendix \ref{sec:Decay_rate} we estimate the decay rate of scalar fields for different types of the SF potential around the central minimum. Finally, in Appendix \ref{photon_sphere} we use the relations obtained in the main text to establish conditions for the existence of a photon sphere around the SSS configuration with SF.

\section{Model Formulation}\label{notations}
We consider $n$ real scalar fields $\Phi\equiv\{\phi^1,...,\phi^n\}$ defined on a $D$ -- dimensional ($D\ge 4$) pseudo-Riemannian manifold $M$, $\Phi: M\to \mt$, where target space $\mt$ is an $n$ -dimensional  Riemannian manifold. 

 The system is described by the action\footnote{We use units $G=c=1$.}
\begin{equation}\label{action_1}
S=\int d^{D} x \, \sqrt{|g|} \,\left[-\frac{1}{16\pi}R+\mathcal{L}\right],~~D\ge 4,    
\end{equation}
with  a sigma-model Lagrangian 
\begin{equation}\label{Lagrangian_1}
\mathcal{L}=\frac{1}{2} \gamma_{ab}(\Phi)\partial_\mu\phi^a\partial^\mu\phi^b-V(\Phi) .
\end{equation}

Throughout the paper, we use the following assumptions:

\begin{enumerate}
\renewcommand{\labelenumi}{(\alph{enumi})}
\item The target space $\mt$ is a complete Riemannian manifold equipped with a positive definite metric tensor\footnote{Latin indexes $a,b,\dots$ run from 1 to $n$.} $\hat\gamma(\Phi)\equiv \{\gamma_{ab}(\Phi)\} \equiv \{\gamma_{ba}(\Phi)\} $. 
Metric $\hat\gamma(\Phi)$, potential $V(\Phi)$ and their derivatives are smooth functions in $\mt$.
\item Zero point $\mathbf{0}\in \mt$ and 
\eq{\label{V_min1}
V(\mathbf{0})=0,~~\nabla_a V(\Phi)|_{\Phi=\mathbf{0}}=0\,. }
\item The SF potential is nonnegative
\eq{\label{V_min}
V(\Phi)\geq 0\,,   \quad \Phi \in \mt \,.  
}
We do not exclude the existence of multiple  minima and/or the case of massless potentials.
\item We consider asymptotically flat SSS systems, assuming that at infinity, the SF components  (along with their derivatives) fall off rapidly enough to preserve the behavior of the Schwarzschild–Tangherlini metric \cite{Tangherlini1963}, which generalizes 
the  Schwarzschild solution ($D=4$) to an arbitrary spacetime dimension $D\ge 4$. 
\end{enumerate}

We shall also need an upper estimate for   $V$. It is convenient to write  this in terms of the target space geometry.

Denote $\Xi(t)\equiv\{\xi^1,...,\xi^n\}:  [t_1,t_2]\to \mt$  a piecewise-smooth curve  in $\mt$, which connects points  $\Xi(t_1)=\Theta_1$ and $\Xi(t_2)=\Theta_2$.  We remind that the length of the curve $\Xi$ between $\Theta_1,\Theta_2$  is defined in a standard way as 
\eq{\label{Length_l_gamma}
\ell_\gamma(\Theta_1,\Theta_2,\Xi)=\int_{t_1}^{t_2}\left[\, \gamma_{ab}(\Xi(t)) \frac{d\xi^a}{dt}\frac{d\xi^b}{dt}\right]^{\frac{1}{2}}dt,
}
 independent of the  regular choices of parameter $t$ on $\Xi$. The Riemannian distance between $\Theta_1,\Theta_2$  is then 
\eq{\label{d_gamma}
d_{\gamma}(\Theta_1,\Theta_2)=\inf\limits_{\Xi}  \ell_\gamma(\Theta_1,\Theta_2,\Xi),
}
where the infimum is taken over all possible piecewise smooth curves $\Xi$ connecting  $\Theta_1,\Theta_2$.

We shall consider potentials such that
\begin{equation}\label{V_upper}
    |V(\Theta)|\leq C_0 \exp\left\{\kappa \left[\,d_\gamma(\Theta,{\mathbf 0})\right]^p\right\}\,,\,  1\le p<2\,. 
\end{equation}
 $0<\kappa<\infty$, $0<C_0<\infty$. 

 Note that in case of the Euclidean  metric $\gamma_{ab}$, we have $d_\gamma(\Theta,0)\equiv\|\Theta\|_E$, where $\|\cdot\|_E$ is the usual Euclidean norm in $\mathbb{R}^n$.

\section{Sigma model in static spherically symmetric configurations}\label{sigma_mod}
\subsection{Basic equations}
 We use the metric of the SSS $D$-dimensional space-time in the form
\begin{equation}
    ds^2=  e^\alpha dt^2-e^\beta dr^2-r^2d\Omega_{D-2}^2,
    \label{SS-metric}
\end{equation}
\noindent where 
$ \alpha =\alpha (r),\, \beta =\beta (r)$, and $d\Omega_{D-2}^{2}$ is the metric of the  $(D-2)$ -- dimensional unit sphere.

The independent Einstein equations in the case of  SSS configuration corresponding to (\ref{action_1})   take the form
\begin{align}
re^{-\beta} \frac{d\beta}{dr}-q(e^{-\beta}-1)
=\frac{8\pi}{q+1} r^2  \left[ 
 e^{-\beta}K+ 2V(\Phi)\right]
    \label{Ein_1-0+} \,,
\end{align}
\begin{align} 
{re^{-\beta} \frac{d\alpha}{dr}+q(e^{-\beta}-1}) = \frac{8\pi r^2}{q+1}  \left[e^{-\beta}K-  2V(\Phi)\right]
 \label{Ein_2-0+}\,,
\end{align}
 where $q=D-3$ and
\begin{equation}
K(r) =\gamma_{ab} \frac{d\phi^a}{dr}\frac{d\phi^b}{dr}   \,, \label{K_pos_def}
\end{equation}
Equations (\ref{Ein_1-0+}), (\ref{Ein_2-0+}) must be considered jointly with scalar field equations corresponding to action (\ref{action_1}):
\begin{align} \nonumber
   \frac{1}{r^{q+1}}\frac{d}{d r}\left[r^{q+1} e^{\frac{\alpha-\beta}{2}}  \gamma_{ab}\frac{d\phi^b}{dr}\right]=& \frac{1}{2}e^{\frac{\alpha-\beta}{2}}\frac{\partial \gamma_{cb}}{\partial\phi^a}\frac{d\phi^c}{dr}\frac{d\phi^b}{dr}+\\
   +&e^{\frac{\alpha+\beta}{2}}  V'_a(\Phi),
\label{eq_field+}
\end{align}
where  $V'_a\equiv \partial V/ \partial \phi^a$.

According to prescription (d), 
corresponding requirements for $\alpha$ and $\beta$ ($r\to\infty$) are  
\begin{align}\label{asympt_metric}
\lim\limits_{r\to\infty}    r^{q} (e^\alpha -1) =\lim\limits_{r\to\infty} r^{q} (e^{-\beta}-1)=- {r_g},
\end{align}
where $r_g=16\pi M/(q+1)\Omega_{D-2}$ is the gravitational radius, $M$ is the  ADM configuration mass,  and $\Omega_{D-2}$ is the volume  of the  ($D-2$) dimensional unit sphere. 
We can guarantee condition (\ref{asympt_metric}), if  SF decays fast enough at spatial infinity. More specifically, we can assume
\begin{equation}\label{Dphi-inf}
\left|{\phi_i}\right|   \le \frac{C}{ r^{q}},~i=1,\ldots,n; \quad r\to\infty\,.
\end{equation}
In fact, the SF can decay faster than (\ref{Dphi-inf}), however, it cannot decay more slowly (see Appendix \ref{sec:Decay_rate}). The precise asymptotic behavior of $\Phi(r)$ depends on the leading terms in the expansion of $V(\Phi)$ near zero, but (\ref{Dphi-inf}) is always fulfilled. For example, if the expansion of $V(\Phi)$  begins with second order terms, then $\Phi(r)$ decays exponentially (see Appendix \ref{flat-flat}).
Under assumption (d), from  equation  (\ref{Ein_2-0+}), we also have 
\begin{equation}\label{alpha'}
\lim\limits_{r\to\infty}\left|r\frac{d\alpha}{d r}\right|=0\,.
\end{equation}

\subsection{Qualitative properties of the solutions}\label{subsec: qualitative}
Our consideration involves two steps. In the first step we consider the solution on $(r_0, r_1]$,  ($0<r_0<r_1$) and prove that it can be extended to the left half-neighborhood of $r_0$. With this aim it is sufficient to prove that the right-hand sides of the system (\ref{eq:fun_Y})--(\ref{eq:fun_X1}) are bounded for $r\to r_0$.

In the second step we extend regular solutions for all $r>0$ by the continuous induction and  consider limit $r_0\to 0$.

We say that the solution  is trivial, if  $\Phi(r)\equiv \mathbf{0}$ is satisfied on some open interval of the radial variable. In this case the solution  of the system (\ref{eq:eq_X})--(\ref{eq:phi}) with conditions (\ref{asympt_metric}) leads to the Schwarzschild--Tangherlini  solution.
In what follows, we exclude trivial cases. 
Assuming a non-trivial solution (i.e., at least one component $\phi^a \not\equiv 0$) for sufficiently large values of $r$, we move from large $r$ to smaller ones

 Define
\eq{\label{eq:XY_def}
X=e^{(\alpha+\beta)/2}>0\,,\quad Y=r^{q} e^{(\alpha-\beta)/2}>0,~r>0,
}
and 
\eq{\label{eq:Z_def}
 Z_a=-r Y\gamma_{ab}\frac{d\phi^b}{dr}\,,\quad Z^a=\gamma^{ab}Z_b=-{rY}\frac{d\phi^a}{dr}\,.
}
In terms of these new variables, the  conditions (\ref{asympt_metric}) can be  written as
\eq{
\label{eq:XY_asympt}
\lim\limits_{r\to\infty}\left[r^q(X-1)\right]=0,
~
\lim\limits_{r\to\infty}\left[Y-r^q\right]=-r_g .
}
The system of  equations (\ref{Ein_1-0+})--(\ref{eq_field+}) can be reduced to 
\eq{
\label{eq:eq_X}
\frac{dX}{dr}= \frac{8\pi r}{q+1}X K,
}
\eq{
\label{eq:fun_Y}
\frac{dY}{dr}=q r^{q-1} X\left[1-\frac{16\pi r^{2}}{q(q+1)}  V(\Phi)\right] \,,
}
\eq{\label{eq:fun_Z}
 \frac{dZ_a}{d r}=-\frac{1}{2rY}      \frac{\partial \gamma_{cd}}{\partial\phi^a} {Z^c}  {Z^d} -r^{q+1} X V'_a(\Phi),
}
and 
\eq{\label{eq:phi}
\frac{d\phi^a}{dr}=-\frac{Z^a}{r Y} \,.
}
Using (\ref{eq:Z_def}), equation  (\ref{eq:eq_X}) yields
\eq{\label{eq:fun_X1}
\frac{dX}{dr}=\frac{8\pi   X \gamma_{ab}Z^a Z^b}{(q+1)r Y^2}\,.
}
We introduce the auxiliary function $F(r)$
\eq{\label{eq:F_function}
F(r)\equiv \frac{1}{q r^{q-1} X}\frac{d}{dr}\left[X Y\right].
}
Also, using (\ref{eq:XY_def}) 
\eq{\label{eq:F_}
F(r)\equiv \frac{1}{q r^{q-1} X}\frac{d}{dr}\left(r^{q} e^\alpha\right)=\left[1+\frac{r}{q}\frac{d \alpha}{dr}\right]e^{(\alpha-\beta)/2}\,.
}
Then, on account of (\ref{asympt_metric}) and (\ref{alpha'}) we have 
\begin{equation}\label{F_infty}
    \lim\limits_{r\to \infty}F(r)=1\,.
\end{equation}


{\it Lemma} 1.
Let  $X,Y,Z_a\in C^{(1)}(r_0,\infty)$ and $\Phi\in C^{(2)}(r_0,\infty)$, $X(r)$, $Y(r)$  satisfy (\ref{eq:fun_Y})--(\ref{eq:fun_X1}) on $(r_0,\infty)$, $r_0> 0$. Let conditions (\ref{eq:XY_asympt})  be satisfied. 

Then  $\forall r>r_0 $ for a nontrivial $\Phi(r)$.   
\begin{enumerate}[label=(\roman*)]
    \item  $ X<1$;
    \item $ F'(r)\leq 0, ~~F(r)> 1$;
    \item $\exists\, \eta>0:~ Y(r)\ge\eta$;
    \item $\exists\, \zeta_0$:  such that for $r\to r_0$ we have\\ $\gamma_{ab}Z^aZ^b \ge \zeta_0> 0$. 
 \end{enumerate}
 
{\it Proof.} 
(i) The RHS of (\ref{eq:fun_X1}) is non-negative, hence $X$ is a non-decreasing function. Using (\ref{eq:XY_asympt}), we have $X\to 1$ as $r\to\infty$.  To get the  strict inequality we need to rule out the case $X=1$.   Let, on the contrary, $X(r_1)=1$ for some $r=r_1<\infty$, then $X(r)\equiv 1$, $\forall r\ge r_1> r_0$. Therefore $X'(r)\equiv 0$ for all $r\ge r_1$. 
The RHS of (\ref{eq:eq_X}) is positive definite, then  we  have that  $\Phi'(r)\equiv \mathbf{0}$ for $r\in [r_1,\infty)$. Therefore the components of $\Phi(r)$ are constant and as $\Phi(r)\to\mathbf{0}$ for $r\to\infty$,  we obtain that $\Phi(r)\equiv \mathbf{0}$ and $Z_a(r)\equiv 0$,  which contradicts the assumption that $\Phi$ is a nontrivial solution. 
This proves (i).\\

(ii) The estimate for $F$ follows from (\ref{eq:eq_X})--(\ref{eq:fun_Y}), which yield 
\eq{\label{eq:F_estimate}
\frac{dF}{dr}=-\frac{32\pi r X(r)V(\Phi)}{q(q+1)} - \frac{8\pi {Y(r)} K(r) }{(q+1)r^{q-1}}\leq 0 .
}
Since $V\geq 0$, $Y>0$, $X>0$ and $K\ge 0$, then $F$ is a monotonically non-increasing function. From (\ref{F_infty}), we have $F(r)\to 1$ as $r\to\infty$. Then $F(r)\ge 1$. If, on the contrary, $F(r)=1$ at some $r_1\ge r_0$, then $F(r)\equiv 1$ for $r>r_1$ and in  (\ref{eq:F_estimate}) $F'(r)\equiv 0$ whence $K\equiv 0$.  This reasoning leads to strict inequality (ii) in case of the non-trivial solution. \\  

(iii) We have from (ii) that
\eq{\label{eq:dXY_aux}
\frac{d}{dr}(XY)=q r^{q-1}X(r)F(r)> q r^{q-1} X(r).
}
On the other hand, from (\ref{eq:fun_Y}), taking into account positivity of $V$,
\begin{equation}
   \frac{dY}{dr}\le q r^{q-1} X <  \frac{d}{dr}(XY)\,,
\end{equation}
whence 
\eq{
   \frac{d}{dr}[(1-X)Y]<0,
}
showing that function $(1-X)Y$ is  monotonically decreasing. Then  
\begin{equation}\label{lower_bound_Y}\eta \triangleq
(1-X(r_1))Y(r_1)<(1-X(r))Y(r)<Y(r),
   \end{equation}
for $r<r_1$.

Note that from (\ref{eq:XY_asympt}), we have $Y(r)\sim r^{q}$ for large $r$ and one can find $r_1>r_0$ (at least, sufficiently large) such that $Y(r)\ge\eta_0 >0\,, \forall r>r_1$. 
By putting $\eta=[1-X(r_1)]\eta_0$ we have statement (iii) of this  {\it Lemma}.\\

  (iv) Consider $r\in (r_0,r_1]$, $r_1>r_0$.  From   (\ref{eq:F_function})  and using (\ref{eq:eq_X}),   (\ref{eq:fun_Y}), we have  
  \begin{align}\label{K>0}
 &\frac{8\pi }{q(q+1)}{r^{2-q} Y(r)}K(r) =  F(r)-X(r)\left[1-\frac{16\pi r^2 V}{q(q+1)}\right]\nonumber \\ \ge &F(r)-X(r)\ge \chi_0>0;\quad \chi_0=F(r_1)-X(r_1)\,,
  \end{align} 
because $F>1$ is decreasing and $X<1$ is increasing. Using  (\ref{lower_bound_Y}), (\ref{K>0}), 
\begin{equation}\label{Z>0}
\gamma_{ab}Z^aZ^b={r^2Y^2}K\ge \frac{q(q+1)r_0^{q}}{ 8\pi} \eta\chi_0   \triangleq \zeta_0 >0\,.
\end{equation}
  This completes the proof of  statement (iv). 
  $\blacksquare$\\

{\it Lemma} 2. Let  $X,~Y,~Z_a\in C^{(1)}(r_0,r_1]$ and $\Phi\in C^{(2)}(r_0,r_1]$,    satisfy  (\ref{eq:fun_Y})--(\ref{eq:fun_X1}) on $(r_0,r_1]$, $r_1>r_0> 0$. 
Let condition (\ref{V_upper}) is fulfilled. 
Then,    for a nontrivial $\Phi(r)$, \\
(i) there are finite limits $X_0>0$, $Y_0>0$:
\eq{ \label{limits_lemma_2+} X_0=\lim\limits_{r\to r_0+0}X(r)<1\,,~
 Y_0=\lim\limits_{r\to r_0+0}Y(r)>0\,, 
}
and finite limits $Z_{a,0}, \phi^a_0$:
\eq{\label{limits_lemma_3+}
\phi^a_0=\lim\limits_{r\to r_0+0}\phi^a(r),  ~Z_{a,0}=\lim\limits_{r\to r_0+0}Z_a(r)\,.
}
(ii) the solution is smoothly extended to the left half-neighborhood of $r_0$.

 {\it Proof.} 
Because $X(r)$ is non-decreasing, the existence of limit $X_0<1$ follows from the Monotone Convergence Theorem. 
The lower estimate  $Y_0>0$ follows from  (iii) of the previous lemma. 

To show that  $Y_0$ is finite, we  shall estimate the RHS of (\ref{eq:fun_Y}) for $r<r_1$. 
From (\ref{eq:eq_X})   follows
\begin{equation}\label{X(r)}
X(r)=X(r_1)\exp\left[ - \frac{8\pi}{q+1} \int\limits_r^{r_1} dt  K(t)\,t \right]  ,
\end{equation}
From Cauchy-Bunyakovsky inequality 
\[ 
\int\limits_r^{r_1}   \sqrt{ K(t)}dt\equiv
\int\limits_r^{r_1} dt \,\, t^{-1/2}\sqrt{t\,K(t)}\le\left[ L(r) \int\limits_r^{r_1} dt\,K(t)t \right]^{1/2},
\]
where $L(r)=\ln(r_1/r)$. 

Note that the left-hand side of this relation is just the length of the curve $\Xi=\Phi(r) \in \mt$ between points $\Theta=\Phi(r)$ and $\Theta_1=\Phi(r_1)$ that cannot be smaller than the  Riemannian distance (\ref{d_gamma}) between $\Theta$ and $\Theta_1$:
\begin{equation}\label{Length_pq}
 \int\limits_r^{r_1} dt \sqrt{K(t)}\equiv   \ell_\gamma(\Theta,\Theta_1,\Phi) \ge  d_{\gamma}(\Theta,\Theta_1) \,. 
\end{equation}
 For $r\ge r_0$,  we have $L(r)\le L(r_0)$, then (\ref{X(r)}), (\ref{Length_pq}) yield
\begin{align}\label{X_est}
   & X(r)\le  X(r_1)\exp\left\{ - \frac{8\pi}{(q+1)L(r_0)} \left[ d_{\gamma}(\Theta,\Theta_1)\right]^2 \right \} \,.
\end{align}
 This enables us to estimate in (\ref{eq:fun_Y}) 
 \begin{equation} \label{B(r)}
B(r)=	 X(r)V(\Phi(r))\,.
\end{equation}
Owing to (\ref{V_upper}), using the triangle inequality $d_\gamma(\Theta,0)\le d_\gamma(\Theta,\Theta_1)+d_\gamma(\Theta_1,0)$, we get
\begin{align}\nonumber 
 &\ln B(r)\le \ln C_0 X(r_1)-  \\ &    - \frac{8\pi}{(q+1)L(r_0)} \left[d_\gamma(\Theta,\Theta_1)\right]^2+\kappa [d_\gamma(\Theta,\Theta_1)+d_\gamma(\Theta_1,0)]^p 
\label{expression}
\end{align}
Evidently, for $p<2$\footnote{As well as for $p=2$.},  this  expression is bounded above (regardless of whether $d_\gamma(\Phi(r),\Phi(r_1))\to\infty$  as $r\to r_0$, or not).
 Thus the right-hand side of (\ref{eq:fun_Y}) is integrable and $Y_0$ in (\ref{limits_lemma_2+})  is bounded.

Now we can estimate $Z^a$.  
Denote
 \begin{align}
 P(r)=\gamma_{ab}Z^a Z^b,~~ \mathcal{E}(r) =P(r)-2r^{q+2}Y(r)B(r),
     \label{E}
 \end{align}
 Using (\ref{eq:fun_Z}) and   (\ref{eq:phi}), we have 
\begin{align*} 
\frac{dP}{dr}=- 2 r^{q+1} X Z^a     V'_a,
    \end{align*}
Using (\ref{eq:fun_Y})--(\ref{eq:fun_X1}), we have  
\begin{equation}\label{equation_E}
    \frac{d\mathcal{E}}{dr}=\mathcal{F}-g(r)\mathcal{E}\,,
\end{equation}
where 
\begin{align*}
\mathcal{F}=&-2r^{q+1}B\left[(q+2)Y+q r^q X\right],~g=\frac{16\pi r^{q+1}}{q+1} \frac{B}{Y}.   
\end{align*}
Here, $B\ge 0$, $X<1$ and $Y$ are uniformly bounded  above  and $Y\ge \eta>0$. Therefore, $\mathcal{F},g$ are bounded and the solution of (\ref{equation_E}) with respect to $\mathcal{E}$ is also bounded
 \begin{align}
 \mathcal{E}(r) & = \mathcal{E}(r_1) \exp\left[\int\limits_r^{r_1} g(x)dx\right]-\nonumber \\ &-\int\limits_r^{r_1}\exp\left[\int\limits_r^{x} g(x')dx'\right]\mathcal{F}(x)dx.
 \label{esteem_Z0}
 \end{align}
Consequently, there exists $\mathcal{E}_0<\infty$:  $\mathcal{E}(r)\le \mathcal{E}_0$ for $r\in (r_0, r_1]$. Then  
\eq{\label{mcl_Z}
P=  \mathcal{E}(r) + &2r^{q+2} Y(r)B(r)\le \mathcal{E}_0+2r_1^{q+2} Y_1 B_0<\infty,
}
which is bounded as $r\to r_0$, i.e., $\exists P_0$: $P(r)\le P_0$. 
Then, we  have estimation for the kinetic term
$ K= P(r)/(rY)^{2}
\le P_0/(\eta r)^{2}$. 

Consider sequence $\{r_n\}\to r_0+0$. By definitions  (\ref{Length_l_gamma}), (\ref{d_gamma}) and using (\ref{Length_pq}), we have ($r_m,r_n\to r_0$, $n,m\to \infty$),
\eq{
&d_{\gamma}(\Phi(r_n),\Phi(r_m))\le  \ell_\gamma(\Phi(r_n),\Phi(r_m),\Phi)=\nonumber\\=&\int_{r_n}^{r_m}  \frac{\sqrt{P(s)}}{s Y}ds\le\frac{\sqrt{P_0}}{\eta}\int_{r_n}^{r_m}\frac{ds}{ s } \le \frac{\sqrt{P_0}}{ r_0\eta} | r_m-r_n |\to 0 .
}
Therefore, $\Phi(r_n)$  is a Cauchy sequence under Riemannian metric $\gamma_{ab}$. As we assumed that $\mt$ is complete, then there is 
a finite limit $\Phi(r_n)\to \Phi_*\in\mt$. In the neighborhood of $\Phi_* $  the topology induced by the Riemannian distance $d_\gamma$ is equivalent to the standard manifold topology,
which locally matches the Euclidean metric.  This implies that $\Phi(r)$ is uniformly bounded for $\forall r\in  (r_0, r_1]$ and $\Phi(r)\to \Phi_0$ for $r\to r_0$.
 
In order to estimate $Z^a$, we note that in view of smoothness and positive definiteness of $\gamma_{ab}$ there exists a uniform constant $\gamma_0 > 0$  such that in the compact neighborhood of $\Phi_*$, we have that there exists some $\gamma_0>0$ such that 
	\begin{equation}
		\sqrt{P(r)}=\sqrt{\gamma_{ab}(\Phi(r))Z^aZ^b} \geq \gamma_0  \left[\sum\limits_{a=1}^{n}(Z^a)^2\right]^{1/2},
	\end{equation}
	for all $r \in (r_0, r_1]$.

  Now we see that the RHS of system (\ref{eq:fun_Y})--(\ref{eq:fun_X1}) are bounded and integrable as functions of $r$, whence we infer that there exist finite limits (\ref{limits_lemma_2+}), (\ref{limits_lemma_3+}) $X_0,Y_0>0,Z_{a,0},\,\phi_0^a$;  RHS   of (\ref{eq:fun_Y})--(\ref{eq:fun_X1})
are Lipschitz-continuous in the neighborhood of this point for $r\to r_0$. Then

  $\bullet$ the solution of (\ref{eq:fun_Y})--(\ref{eq:fun_X1}) exists in the neighborhood of this (regular)  point and 
 
 $\bullet$ assessments (i)--(iv) of  {\it Lemma} 1 are valid on some compact set $[r'_0,r_0]$ that contains inner point $r=r_0$ yielding (\ref{limits_lemma_2+}),(\ref{limits_lemma_3+}).

 The  {\it Lemma} is proved. 
 $\blacksquare$\\

 We arrive at the following statement:\\

{\it Theorem 1. } 
Let  $X,Y,Z_a\in C^{(1)}(r_0,\infty)$ and $\phi\in C^{(2)}(r_0,\infty)$,    $Y(r)>0$  satisfy (\ref{eq:fun_Y})--(\ref{eq:fun_X1}) on $(r_0,\infty)$, $r_0> 0$. Let conditions (\ref{V_upper}) and  (\ref{eq:XY_asympt})  are satisfied. 

Then,  for a nontrivial $\Phi(r)\not\equiv 0$, the solution can be smoothly extended for all $r\in (0,\infty)$, where assessments (i)--(iii) of the {\it Lemma} 1 are preserved.\\

{\it Proof.} 
Taking into account the assessments of {\it Lemmas }1, 2,  the proof of this  statement follows trivially from the continuous induction. 
$\blacksquare$ \\

Thus, under fairly  general conditions, the solution of the problem is regular  on $(0,\infty)$; no ``spherical" singularities occur for all radii $r>0$.

\section{Asymptotic relations for $r\to 0$}\label{Section: asymptotic relations}

{\it Theorem 2}. Let  $X,Y,Z\in C^{(1)}(r_0,\infty)$ and non-trivial  $\Phi\in C^{(2)}(r_0,\infty)$ with   $Y(r)>0$  satisfy (\ref{eq:fun_Y})--(\ref{eq:fun_X1}) on $(r_0,\infty)$, $r_0> 0$. Let conditions (\ref{V_upper}),  (\ref{Dphi-inf}) and (\ref{eq:XY_asympt})  are satisfied and,  in addition
\[p=1,~~ \kappa^2< 32\pi \frac{D-1}{D-2}.\] 

Then the solution of system (\ref{eq:fun_Y})--(\ref{eq:fun_X1}) is smoothly extended on $(0, \infty)$ and the estimates (i)--(iii) are preserved in this interval; there exist finite 
 limits 
   $X_0,$ $Y_0$
\eq{ \label{limits_theorem_2} X_0=\lim\limits_{r\to 0+}X(r)=0\,,~
 Y_0=\lim\limits_{r\to 0+}Y(r)>0\,, 
}
and finite limit\footnote{Components of $\Phi(r)$ can be unbounded for $r\to 0$.} 
\eq{\label{limits_theorem_2+}
 P_0=\lim\limits_{r\to 0+}P(r)> 0.
}

{\it Proof.}  Possibility of extension of the solution on $(0,\infty)$ follows from {\it Lemma} 2. Consider solution on $(0,r_1]$. We can repeat assessments  of {\it Lemma} 1 on this interval, however the considerations of {\it Lemma} 2 must be modified because $L(r)=\ln(r_1/r)$   becomes infinite for $r\to 0$. Correspondingly, we can leave $L(r)$ instead of $L(r_0)$ in (\ref{X_est}), (\ref{expression}) and complete a prefect square. 
Denoting   $\mathcal{D}(r)= r^{q+1}B(r)$;  $\Phi_1\equiv \Phi(r_1)$ and using $\|\Phi\|_\gamma\equiv d_\gamma(\Phi,0)\le\|\Phi_1\|_\gamma+d_\gamma(\Phi,\Phi_1)$, we have 
\begin{align}\label{bound_D}
\ln \mathcal{D}(r)\le & \ln\{ C_0 r_1^{q+1} X(r_1)\}-\nonumber \\&(q+1)L(r) \left[1-\frac {\kappa^2 }{32\pi} \right]+\kappa\|\Phi(r_1)\|_\gamma \,.
     \end{align} 
For $(q+1)(1-\kappa^2/(32\pi))>-1$ the RHS of (\ref{eq:fun_Y}) is integrable, yielding   a sufficient condition 
\eq{
{\kappa^2 }<32\pi \frac{(q+2)}{(q+1)}=32\pi\frac{D-1}{D-2},
}
 that there is a finite limit $Y_0$ in (\ref{limits_lemma_2+}). 

The inequality $Y_0>0$ is the consequence of the  lower bound (iii) from  {\it Lemma} 1. 

Now we can rewrite $\mathcal{F}$ and $g$ from (\ref{equation_E}) in terms of $D=r^{q+1}B$, yielding 
\begin{equation}\label{mcl_F0}
 \mathcal{F}=-2D[(q+2)Y+ q r^q X],~~ g=\frac{16\pi}{q+1}  \frac{D}{Y}.   
\end{equation}
Here, $D$ and  all terms of $\mathcal{F}$ are integrable for $r\to 0$. Applying  (\ref{esteem_Z0})  we get the existence of finite limit  (\ref{limits_theorem_2+}).

Note that  the equality in (\ref{K>0}) can be rewritten as
\begin{equation}\label{E>0}
    \mathcal{E}=\frac{q(q+1)r^q}{8\pi} Y(F-X)>0\,.
\end{equation}
Then using (i) and (ii) of Lemma 1 we have $\mathcal{E}>0$. Moreover, from (\ref{equation_E}) we see that $\mathcal{E}$ is decreasing, that is $\mathcal{E}(r)>\mathcal{E}(r_1)$ 
 then $P(r)\ge \mathcal{E}(r)\ge \mathcal{E}(r_1)$ ($\forall r<r_1$). This justifies strict inequality    (\ref{limits_theorem_2+}) of this theorem.  
$\blacksquare$ \\

It is convenient to introduce 
\eq{
\eta= \frac{8\pi}{q+1} \frac{P_0}{Y_0^2},
}
Then  using (\ref{eq:eq_X}, \ref{eq:XY_def}), we obtain near the singularity 

\begin{equation}\label{asymp_near_0}
\alpha(r)\sim (\eta-q)\ln r\,,\quad \beta(r)\sim (\eta +q)\ln r\,.    
\end{equation}
Thus, near the central singularity, the metric functions possess a universal logarithmic behavior identical to that of a single canonical scalar field without a self-interaction potential \cite{PhysRevD.40.2564} (see \cite{Fisher,JNW} for $D=4$ case).

Although the asymptotic behavior of the individual  components of $\Phi$ and $Z_a$ remains undetermined in general (except in some simple cases, e.g.  $\gamma_{ab}=\delta_{ab}$ \cite{ZhdSt}), we can still obtain behavior of the norm of the tangent vector to the curve $\Phi(r)\in \mathcal{T}$ 
\eq{
\|\Phi'\|_\gamma\simeq\sqrt{\frac{(q+1)\eta}{8\pi}}\frac{1}{r},
}
Consequently, the length along this curve satisfies
\eq{
\ell_\gamma(\Phi(r),\Phi(r_1),\Phi)\simeq  \sqrt{\frac{(q+1)\eta}{8\pi}} \ln\left(\frac{r_1}{r}\right).
}
Thus, these asymptotic behavior is also universal in form.

\section{Conclusions}\label{Discussion}

We have considered asymptotically flat, static, spherically symmetric configurations in $D$-dimensional General Relativity with minimally coupled nonlinear sigma models. For models with complete Riemannian target space $\mathcal{T}$ and a non-negative self-interaction potential that satisfies (\ref{V_upper}), we have proved that nontrivial solutions can be smoothly extended to all $r>0$, excluding spherical-like (shell-like) singularities at finite radii. This generalizes our previous results \cite{ZhdSt} to nonlinear sigma models with nontrivial target-space metrics and removes the requirement of monotonicity of the self-interaction potential. In particular, our results can be applied to non-monotonic self-interacting potentials, such as $V(\phi)\propto\phi^2(\phi^2-a^2)^{2n}$.

Under the additional restrictions of Theorem 2, the solutions have  a central naked singularity at $r=0$ and the metric functions possess the universal logarithmic asymptotic behavior (\ref{asymp_near_0}), similar to the $D$-dimensional generalization of the Fisher–Janis–Newman–Winicour solution. This asymptotic behavior is independent of the form of target space metric and the self-interaction potential, however the the individual components of  scalar fields can  have different asymptotics.

We have also shown that these configurations can possess at most one unstable photon sphere. For partial cases that satisfy conditions of Theorem 2 we found sufficient condition for its existence and a corresponding estimations for photon sphere radius.

Finally, note that our  results are not restricted to configurations with naked singularities. Lemmas 1 and 2 and Theorem 1 can also be applied to the exterior of a spherically symmetric star for $r\geq R$, where $R$ is the stellar radius. In this case, the exterior solution must be matched to an appropriate interior one, for example, within a hydrodynamical model with scalar fields.

\section*{Acknowledgments}
M.~P.~H. is supported in part by National Science Foundation grant PHY-2609905.
O.~S.~S. is supported 
by the Harvard Scholar at Risk Program. V.~I.~Z. is supported by Ministry of Education and
Science of Ukraine (project No. 0124U001660).

\appendix

\section{The decay rate of scalar fields at  infinity} \label{sec:Decay_rate}
\subsection{Upper estimates in a general case}

The aim of this Appendix is to explain  inequality (\ref{Dphi-inf}) concerning the behavior of  $\Phi(r)$ at spatial infinity.  We shall show that this condition is necessary to provide the asymptotic condition (\ref{asympt_metric}).

Using (\ref{asympt_metric}), (\ref{eq:XY_def}), (\ref{B(r)}), for large enough $r$  we have from (\ref{equation_E})  
\begin{equation}
    \frac{d\mathcal{E}}{dr}\le -(q+1)r^{2q+1}V(\Phi)\,,
\end{equation} 
i.d., $\mathcal{E}$ is monotonically decreasing
and, integrating over $[r_0,\infty)$, for $r>r_0$,  
\eq{\label{eq:est_int_V}
\int\limits_{r}^{\infty}s^{2q+1} V(\Phi(s))ds\le \frac{\mathcal{E}(r)}{q+1}\le \frac{\mathcal{E}_0}{q+1},
}
where $\mathcal{E}_0=\mathcal{E}(r_0)$
In the neighborhood of $\Phi=\mathbf{0}$,  we have $\exists C_1>0$: 
\eq{\label{eq:est_inf_gamma}
K\equiv \gamma_{ab}\frac{d\phi^a}{dr}\frac{d\phi^b}{dr}\ge C_1 \|\Phi'\|_E^2\ge C_1 \left(\frac{d\phi^a}{dr}\right)^2,
}
This enables to estimate $|d\phi^a/dr|$ by means of ${K(r)}$.
Using (\ref{E}), (\ref{eq:XY_def}) and monotonicity of $\mathcal{E}(r)$, for $r\to \infty$
\begin{equation*}
    P(r)\approx \mathcal{E}_0+2r^{2q+2}V(\Phi(r)),
\end{equation*}
and  
\begin{align*}
    \sqrt{K(r)}=\frac{\sqrt{P(r)}}{r Y(r)}<&\left[ \frac{4\mathcal{E}_0}{r^{2q+2}}+4V(\Phi(r))\right]^{1/2} \\& <\frac{2\sqrt{\mathcal{E}_0}}{r^{q+1}}+2\sqrt{V(\Phi(r))}\,.
\end{align*}
Then 
\eq{
\int\limits_{r}^{\infty}\sqrt{K(s)}ds < \frac{2\sqrt{\mathcal{E}_0}}{q r^q}+2\int\limits_{r}^{\infty}\sqrt{V}ds,
}
For the second term we apply Cauchy-Bunyakovsky inequality and use (\ref{eq:est_int_V}) 
\begin{align*}
 \int\limits_{r}^{\infty}\sqrt{\frac{s^{2q+1}}{s^{2q+1}}V}ds&\le
\left[\frac{1}{2q r^{2q}}\int\limits_{r}^{\infty}s^{2q+1} V ds\right]^{\frac{1}{2}}\\&\le \frac{\sqrt{\mathcal{E}_0}}{\sqrt{2q(q+1)}r^q}\,.
\end{align*}
At last
\eq{
\int\limits_{r}^{\infty}\sqrt{K(s)}ds <  \frac{2\sqrt{\mathcal{E}_0}}{r^q}\left[\frac{1}{q}+\frac{1}{\sqrt{2q(q+1)}}\right],
}
By virtue of (\ref{eq:est_inf_gamma}) this yields   
\eq{
|\phi^a(r)|\le \frac{1}{\sqrt{C_1}}\int\limits_{r}^{\infty}\sqrt{K(s)}ds\le \frac{C_0}{r^q},
}
in accordance with (\ref{Dphi-inf}).

\subsection{Asymptotic relations for special cases}
Here we present more exact asymptotic relations depending on the expansion of the SF potential near the minimum. We will restrict ourselves to extreme cases of asymptotics at infinity, from exponential decay to the power-law case $|\Phi|\sim r^{-q}$.

Let target space metric $\hat\gamma(\Phi)=\hat\gamma(0)+O(\Phi)$ in the neighborhood of $\Phi=0$. Below, we estimate the leading terms of the asymptotic behavior in case of  flat  $D$-dimensional space-time and constant metric tensor $\hat\gamma(0)$. Typically the next order terms in $\hat\gamma(\Phi)$ do not contribute into the leading terms of the corresponding asymptotics. The same can be said about contribution of the Tangherlini metric \cite{Tangherlini1963} for $D>4$. The only exception is the  massive SF with  $D=4$; in this case, taking into account the Schwarzschild metric changes the power factor, but does not change the exponential decay of the field \cite{Asanov1968,Asanov1974,Rowan1976}, see also \cite{St-Zhd_2019UJP}.

To construct the asymptotic solutions at spatial infinity, we consider the eigenvalues and eigenvectors of the Hessian operator $A^{a}_{b}=\left.\gamma^{ac}\partial_{c}\partial_bV(\Phi)\right|_{\Phi=\mathbf{0}}$. In general, this operator can have both positive and zero eigenvalues, corresponding to massive and massless fields, respectively. Different scalar field components $\phi^a$ may have different asymptotic behaviors or their combinations. For simplicity, we consider only two cases that correspond to  purely massive and purely massless fields.

\subsubsection{Massive scalar fields}\label{flat-flat} 
Consider  the field equation (\ref{eq_field+}) in flat manifold $\mt$ with positive definite metric tensor $\hat \gamma(0)$ and flat $D$-dimensional space-time ($e^\alpha=e^\beta=1$)
\begin{align} 
   \frac{d}{d r}\left[r^{q+1}   \hat\gamma \frac{d\Phi}{dr}\right]= r^{q+1}   \hat m\Phi,
\label{eq_field+flat1}
\end{align}
where  $\hat m$ - is a positive definite symmetric matrix (a generalization of the mass squared such that  
$V(\Phi)=\frac12 \Phi^T\hat m \,\Phi$). Multiplying by $(\hat\gamma)^{-1}$, we have 
\begin{align} 
   \frac{d}{d r}\left[r^{q+1}   \frac{d\Phi}{dr}\right]= r^{q+1}   \hat M\Phi,
\label{eq_field+flat2}
\end{align}
where the matrix $\hat M=(\hat\gamma)^{-1}\hat m$ is similar to a positive definite matrix and therefore has positive eigenvalues $\mu^2_{(p)>0}$ corresponding to  eigenvectors $V_{(p)}$, i.e. $\hat T V_{(p)}=\mu^2_{(p)} V_{(p)}$. Putting 
$\Phi=\sum\limits_{p=1}^n c_p(r)V_p $, we split the system (\ref{eq_field+flat2}) into  linear equations  for the  individual $c_p(r)$: 
\begin{equation} 
   \frac{d}{d r}\left[r^{q+1}   \frac{dc_p}{dr}\right]= r^{q+1}   \mu^2_{(p)} c_p.
\label{eq_field+c_p}
\end{equation}
that lead  to modified Bessel equations. Taking into account condition $\Phi\to 0$ for $r\to\infty$, we discard the exponentially growing terms yielding for $q>1$ ($D>4$)
\begin{equation}\label{eq_field+solution}
    \Phi=r^{-q/2}\sum\limits_{p=1}^n Q_p V_{(p)} K_{\nu}(\mu_{(p)}r)\,,\quad \nu=q/2\,,
\end{equation}
where $Q_p$ are constants and  $K_\nu$ are the modified Hankel functions having asymptotic behavior $K_\nu(x)\sim e^{-x}/\sqrt{x}$. 
This function decays even faster than is required by (\ref{Dphi-inf}).  Hovever, the case $q=1$ is somewhat different (see below).

In the Schwarzschild-Tangherlini space-time, the field equation  equation (\ref{eq_field+c_p}) must be modified as follows
\begin{equation} 
   \frac{d}{d r}\left[r^{q+1} \left[1-\left(\frac{r_g}{r}\right)^p\right]  \frac{dc_p}{dr}\right]= r^{q+1}   \mu^2_{(p)} c_p.
\label{eq_field+Tangh}
\end{equation}
Due to the term $\sim r_g/r$ in the metric in case of $p=1$ ($D=4$), the power law factor must be corrected \cite{Asanov1968,Asanov1974,Rowan1976,St-Zhd_2019UJP, ZhdSt} (unlike the case $p>1$):
\begin{equation}
     c_p(r)\sim \frac{Q_p}{r^{1+\kappa}} e^{-\mu_{(p)}r}\,,\quad  \kappa=\frac12r_g\mu_{(p)}
\end{equation}
In case of $p>1$ ($D>4$) the correction due to the term $\sim (r_g/r)^q$ does not contribute to the leading order of (\ref{eq_field+solution}). 

In the case of variable 
$ \hat\gamma(\Phi)=\hat\gamma(0)+O(\Phi)$, it is easy to see that corresponding corrections in (\ref{eq_field+}) contribute to higher order terms  of asymptotic expansion of solution of (\ref{eq_field+}), not to the leading terms. The same can be said concerning the higher order corrections to the potential in (\ref{eq_field+flat1}). In a whole, for massive fields we have an exponential decay at infinity. Of course, condition (\ref{Dphi-inf}) is not violated. 

\subsubsection{Massless scalar fields}
Let us consider another case, when the SF potential is a positive-definite homogeneous function of degree $2m$, where $m > 1$  is an integer.  Correspondingly, ${\mathbf D}V$ of (\ref{eq_field+nonlin-flat-flat}) is a homogeneous function of degree $2m-1$. 
In order to estimate the behavior of $\Phi(r)$ at infinity, 
similarly to subsection (\ref{flat-flat}), we start from the field equation (\ref{eq_field+}) for vector-function $\Phi$ in flat manifold $\mt$ with symmetric constant  $\hat \gamma=\hat \gamma(0)$ and flat $D$-dimensional space-time metric. In this case 
\begin{align} 
    \frac{d}{d r}\left[r^{q+1}     \frac{d\Phi}{dr}\right]=  r^{q+1} (\hat\gamma)^{-1} {\mathbf D}V(\Phi)\,, 
\label{eq_field+nonlin-flat-flat}
\end{align}
where 
\[
{\mathbf D}V\equiv \frac{\partial V}{\partial \Phi}=\left\{\frac{\partial V}{\partial \phi^1},...,\frac{\partial V}{\partial \phi^n}\right\}\,.
\]
 Substitution $\Phi=\Psi/r^k$ with $k=1/(m-1)$ and $t=\ln r$ leads to an autonomous system 
\begin{equation}
\frac{d^2\Psi}{dt^2}+(q-2k)\frac{d\Phi}{dt} +k(k-q)\Psi=(\hat\gamma)^{-1}{\mathbf D}V(\Psi) 
\label{rough system}
\end{equation}
The eigenvalues of the linear part of (\ref{rough system}) are
\[
\lambda_{1,2}=k-\frac12 q \pm \frac12 q.
\]
If $k=1/(m-1)\ne q$ (absence of zero eigenvalues), then (\ref{rough system}) is a rough system; its behavior near the minimum of the SF potential is completely defined by L.H.S. of (\ref{rough system}). This yields the only possible decaying asymptotic solution for $r\to\infty$
\begin{equation}
\phi^a(r)\sim \frac{Q_a}{r^q}\,,  \quad   \frac{d\phi^a}{dr}   \sim -q\frac{Q_a}{r^{q+1}}\,,
\label{asymp-rough}
\end{equation}
in accordance with (\ref{Dphi-inf}). 

Contributions arising from deviations from flat spacetime and from a flat target space do not alter the behavior of the leading terms of expansions for $r \to \infty$. 

The only case, when system (\ref{rough system}) is not rough ($k=q$) corresponds to $D=4$, $m=2$. In this case the nonlinear components in  (\ref{rough system}) may contribute at the same level as the linear terms depending on the detailed structure of $V$. In this case  (\ref{rough system}) is equivalent to
\begin{equation}
  \frac{d\Phi}{dt} =-(\hat\gamma)^{-1}{\mathbf D}V +\frac{d^2\Psi}{dt^2}
\label{non-rough system-2}   
\end{equation}
The leading term of asymptotic expansion for $t\to \infty$ is  defined by 
\begin{equation}
  \frac{d\Phi}{dt} =-(\hat\gamma)^{-1}{\mathbf D}V(\Psi) \,.
 \end{equation}
The RHS. is the polynomial of degree 3 and the solution can be found by putting
 $\Psi= {{\mathbf \xi}}/{\sqrt{t}}
 $, where $\xi$ is a non-zero root of the algebraic equation
$    {\xi}  =2(\hat\gamma)^{-1}{\mathbf D}V(\xi)$. 
Finally, $\Phi(r)\sim r ^{-1}(\ln r)^{-1/2}$. 

\section{Shadows
}\label{photon_sphere}
In this Appendix we derive conditions of existence of a photon sphere in the SSS space-time with metric (\ref{SS-metric}).

The equations of motion of the massless test particles has the first integral
\eq{\label{eq:geod_integral}
g_{\mu\nu}\frac{dx^{\mu}}{d\tau}\frac{dx^{\nu}}{d\tau}=0,
}
where $\tau$ is the affine parameter along geodesics.

Due to the $SO(D-1)$ symmetry, it is sufficient to study the motion of the test particle along a great circle of $S^{D-2}$, i.e. we  can choose the hyperspherical coordinates as
\eq{
\theta_1=\theta_2=\cdots=\theta_{D-3}=\frac{\pi}{2},
}
Together with the two first integrals associated with the Killing vectors $\partial_t$ and $\partial_\phi$
\eq{
e^{\alpha}\frac{dt}{d\tau}=E\,,\quad r^2\frac{d\varphi}{d\tau}=L,
}
the problem of the photon  motion reduces to a one-dimensional problem,
\eq{
\frac{e^{\alpha+\beta}}{L^2}\left(\frac{dr}{d\tau}\right)^2=\frac{E^2}{L^2}-U_{\rm eff},
}
where
\eq{
U_{\rm eff}=\frac{e^{\alpha}}{r^2},
}
Then the equation from the photon orbit radius follows from $U_{\rm eff}'(r)=0$, yielding
\eq{\label{eq_for_r_ph}
r\alpha'-2=0}

{\it Lemma 3.}

i) Solutions of (\ref{Ein_1-0+})--(\ref{eq_field+}) satisfying
(\ref{asympt_metric}) can have at most one photon sphere, which,
if it exists, is unstable.

ii) The radius of any photon sphere is bounded from above:
\eq{
r_{\rm ph}^{q}\le \frac{q+2}{2}\,r_g.
}

iii) If the conditions of {\it Theorem 2} are satisfied, then a photon sphere exists for $\eta>q+2$ and its radius satisfies
\eq{
\frac{Y_0}{2}(\eta-q-2)
\le r_{\rm ph}^{q}
\le \frac{q+2}{2}\,r_g.
}

{\it Proof.}
It is convenient to introduce the  auxiliary function 
\begin{equation}
\label{eq:R_function}
    R(r)=Y(r)\left(2-r\frac{d\alpha}{dr}\right)\,.
\end{equation} 
whose roots coincide with those of (\ref{eq_for_r_ph}).

i) Using (\ref{eq:F_}), we can rewrite (\ref{eq:R_function}) 
as
\eq{\label{eq:R_F}
R=(q+2)Y-qr^qF.
}
Differentiation and using (\ref{eq:eq_X}),
(\ref{eq:fun_Y}), and (\ref{eq:F_estimate}), we obtain
\eq{\label{eq:dR/dr}
\frac{dR}{dr}=2qr^{q-1}X>0,
}
i.e., $R$ is strictly increasing for $r>0$ and can cross the
abscissa axis only once. Therefore, $R(r)$ has at most one root
$r=r_{\rm ph}$, corresponding to the photon sphere radius.
Using
\eq{
\left.\frac{d}{dr}\left[2-r\alpha'\right]\right|_{r=r_{\rm ph}}
=
\frac{2q}{r_{\rm ph}}e^{\beta(r_{\rm ph})}>0,
}
we obtain that $r_{\rm ph}$ is a simple root. From
\eq{
U_{\rm eff}''(r_{\rm ph})
=
-\frac{2qX^2(r_{\rm ph})}{r_{\rm ph}^{4}}<0,
}
i.e. the corresponding photon orbit is unstable.

ii) Integrating (\ref{eq:dR/dr}),  we obtain
\eq{\label{eq:R_int}
R(r)\Big|_{r_1}^{r_2}=2q\int\limits_{r_1}^{r_2}s^{q-1} X(s)ds\le 2(r_2^q-r_1^q). 
}
Then for $r_1=r_{\rm ph}$ and $r_2\to\infty$, with  (\ref{eq:R_F}, \ref{eq:XY_asympt}, \ref{F_infty})
we obtain
\eq{
r_{\rm ph}^q\le\frac{q+2}{2}r_g,
}
where the divergent terms at spatial infinity cancel. The corresponding upper bound coincides with that previously obtained in \cite{Song_2026} for an anisotropic fluid.

iii) Using the asymptotic expansion of (\ref{eq:R_function})
near the singularity (\ref{asymp_near_0}) and at spatial infinity
(\ref{eq:XY_asympt}), (\ref{asympt_metric}), we obtain
\eq{
R(r)\simeq Y_0(q+2-\eta)
}
for $r\to0$, and
\eq{
R(r)\simeq 2r^q\to+\infty
}
as $r\to\infty$. Thus, for $\eta>q+2$, the function $R(r)$ changes sign on $(0,\infty)$, and the photon sphere exists.

To obtain the lower bound, we set $r_1\to0+$ and $r_2=r_{\rm ph}$ 
in (\ref{eq:R_int}), these yields 
\eq{
r_{\rm ph}^{q}
\ge \frac{Y_0}{2}(\eta-q-2).
}
Together with ii), this gives the corresponding two-sided inequality.
$\blacksquare$
\bibliography{similarity.bib}

@ARTICLE{ZhdSt,
       author = {{Zhdanov}, V.~I. and {Stashko}, O.~S.},
        title = "{Static spherically symmetric configurations with $N$ nonlinear scalar fields: Global and asymptotic properties}",
      journal = {\prd},
         year = 2020,
        month = mar,
       volume = {101},
       number = {6},
          eid = {064064},
        pages = {064064},
          doi = {10.1103/PhysRevD.101.064064},
archivePrefix = {arXiv},
       eprint = {1912.00470},
 primaryClass = {gr-qc},
       adsurl = {https://ui.adsabs.harvard.edu/abs/2020PhRvD.101f4064Z}
}

@ARTICLE{St-Zhd_2019UJP,
       author = {Stashko, O.~S. and Zhdanov, V.~I.},
        title = {Black hole mimickers in astrophysical configurations with scalar fields},
      journal = {Ukrainian J. Phys. },
            year = 2019,
        month = nov,
       volume = {64(11)},
        pages = {1078},
          doi={10.15407/ujpe64.11.1078}
   }

@ARTICLE{Fisher,
       author = {{Fisher}, I.~Z.},
        title = "{Scalar mesostatic field with regard for gravitational effects}",
      journal = {Zh. Exp. Theor. Phys., },
         year = 1948,
        pages = {636-640},
           volume={18},
archivePrefix = {arXiv},
       eprint = {gr-qc/9911008},
 primaryClass = {gr-qc},
       adsurl = {https://ui.adsabs.harvard.edu/abs/1999gr.qc....11008F}
}

@ARTICLE{JNW,
       author = {{Janis}, Allen I. and {Newman}, Ezra T. and {Winicour}, Jeffrey},
        title = {Reality of the {S}chwarzschild Singularity},
      journal = {\prl},
         year = 1968,
        month = apr,
       volume = {20},
       number = {16},
        pages = {878-880},
          doi = {10.1103/PhysRevLett.20.878},
       adsurl = {https://ui.adsabs.harvard.edu/abs/1968PhRvL..20..878J}
}

@article{Christodoulou1984,
author = {Christodoulou, Demetrios},
doi = {10.1007/BF01223743},
issn = {0010-3616},
journal = {Communications in Mathematical Physics},
month = {jun},
number = {2},
pages = {171--195},
title = {{Violation of cosmic censorship in the gravitational collapse of a dust cloud}},
url = {http://link.springer.com/10.1007/BF01223743},
volume = {93},
year = {1984}
}

@article{Ori1987,
author = {Ori, Amos and Piran, Tsvi},
doi = {10.1103/PhysRevLett.59.2137},
issn = {0031-9007},
journal = {Phys. Rev. Lett.},
month = {nov},
number = {19},
pages = {2137--2140},
title = {{Naked singularities in self-similar spherical gravitational collapse}},
url = {https://link.aps.org/doi/10.1103/PhysRevLett.59.2137},
volume = {59},
year = {1987}
}

@article{Joshi1993,
author = {Joshi, P. S. and Dwivedi, I. H.},
doi = {10.1103/PhysRevD.47.5357},
issn = {0556-2821},
journal = {Phys. Rev. D},
month = {jun},
number = {12},
pages = {5357--5369},
title = {Naked singularities in spherically symmetric inhomogeneous {T}olman-{B}ondi dust cloud collapse},
url = {https://link.aps.org/doi/10.1103/PhysRevD.47.5357},
volume = {47},
year = {1993}
}

@article{Bekenstein1972,
author = {Bekenstein, Jacob D.},
doi = {10.1103/PhysRevLett.28.452},
issn = {0031-9007},
journal = {Phys. Rev. Lett.},
month = {feb},
number = {7},
pages = {452--455},
title = {Transcendence of the Law of Baryon-Number Conservation in Black-Hole Physics},
url = {https://link.aps.org/doi/10.1103/PhysRevLett.28.452},
volume = {28},
year = {1972}
}

@article{Bekenstein1972a,
author = {Bekenstein, Jacob D.},
doi = {10.1103/PhysRevD.5.1239},
issn = {05562821},
journal = {Phys. Rev. D},
number = {6},
pages = {1239--1246},
title = {{Nonexistence of baryon number for static black holes}},
volume = {5},
year = {1972}
}

@article{Penrose1965,
author = {Penrose, Roger},
doi = {10.1103/PhysRevLett.14.57},
issn = {0031-9007},
journal = {Phys. Rev. Lett.},
month = {jan},
number = {3},
pages = {57--59},
title = {Gravitational Collapse and Space-Time Singularities},
url = {https://link.aps.org/doi/10.1103/PhysRevLett.14.57},
volume = {14},
year = {1965}
}

@article{Penrose2002,
author = {Penrose, R.},
doi = {10.1023/A:1016578408204},
issn = {0001-7701},
journal = {General Relativity and Gravitation},
month = {jul},
number = {7},
pages = {1141--1165},
title = {Gravitational Collapse: The Role of General Relativity},
url = {http://link.springer.com/10.1023/A:1016578408204},
volume = {34},
year = {2002}
}

@article{Ong2020,
author = {Ong, Yen Chin},
doi = {10.1142/S0217751X20300070},
issn = {0217-751X},
journal = {International Journal of Modern Physics A},
month = {may},
number = {14},
pages = {2030007},
title = {{Space–time singularities and cosmic censorship conjecture: A Review with some thoughts}},
url = {https://www.worldscientific.com/doi/abs/10.1142/S0217751X20300070},
volume = {35},
year = {2020}
}

@Article{Asanov1974,
author={Asanov, R. A.},
title={Point source of massive scalar field in gravitational theory},
journal={Theoretical and Mathematical Physics},
year={1974},
month={Jul},
day={01},
volume={20},
number={1},
pages={667-670},
issn={1573-9333},
doi={10.1007/BF01038757},
url={https://doi.org/10.1007/BF01038757}
}

@article{Doneva_2020,
   title={No-hair theorems for noncanonical self-gravitating static multiple scalar fields},
   volume={102},
   ISSN={2470-0029},
   url={http://dx.doi.org/10.1103/PhysRevD.102.084055},
   DOI={10.1103/physrevd.102.084055},
   number={8},
   journal={Phys. Rev. D},
   publisher={American Physical Society (APS)},
   author={Doneva, Daniela D. and Yazadjiev, Stoytcho S.},
   year={2020},
   month={Oct}
}

@article{Wyman,
  title = {Static spherically symmetric scalar fields in general relativity},
  author = {Wyman, Max},
  journal = {Phys. Rev. D},
  volume = {24},
  issue = {4},
  pages = {839--841},
  numpages = {0},
  year = {1981},
  month = {Aug},
  publisher = {American Physical Society},
  doi = {10.1103/PhysRevD.24.839},
  url = {https://link.aps.org/doi/10.1103/PhysRevD.24.839}
}

@article{Virbhadra_1997,
   title={{Janis–Newman–Winicour and Wyman solutions are the same}},
   volume={12},
   ISSN={1793-656X},
   url={http://dx.doi.org/10.1142/S0217751X97002577},
   DOI={10.1142/s0217751x97002577},
   number={27},
   journal={International Journal of Modern Physics A},
   publisher={World Scientific Pub Co Pte Lt},
   author={Virbhadra, K. S.},
   year={1997},
   month={Oct},
   pages={4831–4835}
}

@ARTICLE{Strongly2021,
       author = {{Stashko}, Oleksandr and {Zhdanov}, Valery I.},
        title = {Singularities in Static Spherically Symmetric Configurations of General Relativity with Strongly Nonlinear Scalar Fields},
      journal = {Galaxies},
         year = 2021,
        month = oct,
       volume = {9},
       number = {4},
        pages = {72},
          doi = {10.3390/galaxies9040072},
archivePrefix = {arXiv},
       eprint = {2109.01931},
 primaryClass = {gr-qc},
       adsurl = {https://ui.adsabs.harvard.edu/abs/2021Galax...9...72S}
}

@ARTICLE{Wald1980,
       author = {{Wald}, Robert M.},
        title = "{Dynamics in nonglobally hyperbolic, static space-times}",
      journal = {Journal of Mathematical Physics},
         year = 1980,
        month = dec,
       volume = {21},
       number = {12},
        pages = {2802-2805},
          doi = {10.1063/1.524403},
       adsurl = {https://ui.adsabs.harvard.edu/abs/1980JMP....21.2802W}
}

@article{Who_afraids,
  title = "{Who's afraid of naked singularities? Probing timelike singularities with finite energy waves}",
  author = {Ishibashi, Akihiro and Hosoya, Akio},
  journal = {Phys. Rev. D},
  volume = {60},
  issue = {10},
  pages = {104028},
  numpages = {12},
  year = {1999},
  month = {Oct},
  publisher = {American Physical Society},
  doi = {10.1103/PhysRevD.60.104028},
  url = {https://link.aps.org/doi/10.1103/PhysRevD.60.104028}
}

@ARTICLE{Asanov1968,
       author = {{Asanov}, R.~A.},
        title = {Static Scalar and Electric Fields in {E}instein's Theory of Relativity},
      journal = {Soviet Journal of Experimental and Theoretical Physics},
         year = 1968,
        month = feb,
       volume = {26},
        pages = {424},
       adsurl = {https://ui.adsabs.harvard.edu/abs/1968JETP...26..424A}
}

@ARTICLE{Rowan1976,
       author = {{Rowan}, D.~J. and {Stephenson}, G.},
        title = "{The massive scalar meson field in a Schwarzschild background space}",
      journal = {Journal of Physics A Mathematical General},
         year = 1976,
        month = aug,
       volume = {9},
       number = {8},
        pages = {1261-1265},
          doi = {10.1088/0305-4470/9/8/014},
       adsurl = {https://ui.adsabs.harvard.edu/abs/1976JPhA....9.1261R}
}

@ARTICLE{Ishibashi-Wald2003,
       author = {{Ishibashi}, Akihiro and {Wald}, Robert M.},
        title = "{Dynamics in non-globally-hyperbolic static spacetimes: II. General analysis of prescriptions for dynamics}",
      journal = {Classical and Quantum Gravity},
         year = 2003,
        month = aug,
       volume = {20},
       number = {16},
        pages = {3815-3826},
          doi = {10.1088/0264-9381/20/16/318},
archivePrefix = {arXiv},
       eprint = {gr-qc/0305012},
 primaryClass = {gr-qc},
       adsurl = {https://ui.adsabs.harvard.edu/abs/2003CQGra..20.3815I}
}

@article{Penrose:1969pc,
    author = "Penrose, R.",
    title = "{Gravitational collapse: The role of general relativity}",
    doi = "10.1023/A:1016578408204",
    journal = "Riv. Nuovo Cim.",
    volume = "1",
    pages = "252--276",
    year = "1969"
}

@article{Heusler:1992ss,
    author = "Heusler, M.",
    title = "{A No hair theorem for selfgravitating nonlinear sigma models}",
    doi = "10.1063/1.529899",
    journal = "J. Math. Phys.",
    volume = "33",
    pages = "3497--3502",
    year = "1992"
}

@article{Pereira:2026qpr,
    author = "Pereira, David S. and Lobo, Francisco S. N. and Mimoso, Jos{\'e} Pedro",
    title = "{A Buchdahl solution generator for Einstein-sigma gravity and multiscalar-tensor gravity}",
    eprint = "2608.20180",
    journal = {},
    archivePrefix = "arXiv",
    primaryClass = "gr-qc",
    month = "8",
    year = "2026"
}

@article{Brax:2023now,
    author = "Brax, Philippe and Burgess, C. P. and Quevedo, F.",
    title = "{Lifting Klein-Gordon/Einstein solutions to general nonlinear sigma-models: the wormhole example}",
    eprint = "2308.12004",
    archivePrefix = "arXiv",
    primaryClass = "hep-th",
    doi = "10.1007/JHEP02(2024)130",
    journal = "JHEP",
    volume = "02",
    pages = "130",
    year = "2024"
}

@article{Machet:2025vzt,
    author = "Machet, Ludovico and Arest{\'e} Sal{\'o}, Llibert",
    title = "{Scalar fields from nonlinear sigma models on black hole spacetimes}",
    eprint = "2508.18362",
    archivePrefix = "arXiv",
    primaryClass = "gr-qc",
    doi = "10.1103/klxz-fgmt",
    journal = "Phys. Rev. D",
    volume = "112",
    number = "8",
    pages = "084047",
    year = "2025"
}

@ARTICLE{Herdeiro-Radu2015IJMPD,
       author = {{Herdeiro}, Carlos A.~R. and {Radu}, Eugen},
        title = "{Asymptotically flat black holes with scalar hair: A review}",
      journal = {International Journal of Modern Physics D},
         year = 2015,
        month = jun,
       volume = {24},
       number = {9},
          eid = {1542014-219},
        pages = {1542014-219},
          doi = {10.1142/S0218271815420146},
archivePrefix = {arXiv},
       eprint = {1504.08209},
 primaryClass = {gr-qc},
       adsurl = {https://ui.adsabs.harvard.edu/abs/2015IJMPD..2442014H}
}

@ARTICLE{Doneva2025IJMPD,
       author = {{Yazadjiev}, Stoytcho S. and {Doneva}, Daniela D.},
        title = "{No-hair theorems in general relativity and scalar-tensor theories}",
      journal = {International Journal of Modern Physics D},
         year = 2025,
        month = jul,
       volume = {34},
       number = {9},
          eid = {2530004-346},
        pages = {2530004-346},
          doi = {10.1142/S0218271825300046},
archivePrefix = {arXiv},
       eprint = {2505.01038},
 primaryClass = {gr-qc},
       adsurl = {https://ui.adsabs.harvard.edu/abs/2025IJMPD..3430004Y},
}

@BOOK{Hawking-Ellis1975,
       author = {{Hawking}, S.~W. and {Ellis}, G.~F.~R.},
        title = "{The Large Scale Structure of Space-Time}",
         year = 1975,
        number = {9},
     
}

@article{Agnese1985,
  title = {Gravitation without black holes},
  author = {Agnese, A. G. and La Camera, M.},
  journal = {Phys. Rev. D},
  volume = {31},
  issue = {6},
  pages = {1280--1286},
  numpages = {0},
  year = {1985},
  month = {Mar},
  publisher = {American Physical Society},
  doi = {10.1103/PhysRevD.31.1280},
  url = {https://link.aps.org/doi/10.1103/PhysRevD.31.1280}
}

@article{Tangherlini1963,
  title = {Schwarzschild field in n dimensions and the dimensionality of space problem},
  author = {Tangherlini, F. R.},
  journal = {Nuovo Cimento},
  volume = {27},
   pages = {636–651},
    year = {1963},
  doi = {10.1007/BF02784569},   
}

@article{PhysRevD.102.084032,
  title = {Rotating tensor-multiscalar black holes with two scalars},
  author = {Collodel, Lucas G. and Doneva, Daniela D. and Yazadjiev, Stoytcho S.},
  journal = {Phys. Rev. D},
  volume = {102},
  issue = {8},
  pages = {084032},
  numpages = {12},
  year = {2020},
  month = {Oct},
  publisher = {American Physical Society},
  doi = {10.1103/PhysRevD.102.084032},
  url = {https://link.aps.org/doi/10.1103/PhysRevD.102.084032}
}

@article{PhysRevD.104.064045,
  title = {Nonlinear stability of soliton solutions for massive tensor-multiscalar theories},
  author = {Falcone, Riccardo and Doneva, Daniela D. and Kokkotas, Kostas D. and Yazadjiev, Stoytcho S.},
  journal = {Phys. Rev. D},
  volume = {104},
  issue = {6},
  pages = {064045},
  numpages = {10},
  year = {2021},
  month = {Sep},
  publisher = {American Physical Society},
  doi = {10.1103/PhysRevD.104.064045},
  url = {https://link.aps.org/doi/10.1103/PhysRevD.104.064045}
}

@article{PhysRevLett.109.251102,
  title = {Surface Singularities in Eddington-Inspired Born-Infeld Gravity},
  author = {Pani, Paolo and Sotiriou, Thomas P.},
  journal = {Phys. Rev. Lett.},
  volume = {109},
  issue = {25},
  pages = {251102},
  numpages = {5},
  year = {2012},
  month = {Dec},
  publisher = {American Physical Society},
  doi = {10.1103/PhysRevLett.109.251102},
  url = {https://link.aps.org/doi/10.1103/PhysRevLett.109.251102}
}

@article{Nagy:2021bha,
    author = "Nagy, Cec{\'\i}lia and Keresztes, Zolt{\'a}n and Gergely, L{\'a}szl{\'o} {\'A}.",
    title = "{Spherically symmetric, static black holes with scalar hair, and naked singularities in nonminimally coupled k-essence}",
    eprint = "2210.00287",
    archivePrefix = "arXiv",
    primaryClass = "gr-qc",
    doi = "10.1103/PhysRevD.103.124056",
    journal = "Phys. Rev. D",
    volume = "103",
    number = "12",
    pages = "124056",
    year = "2021"
}

@article{PhysRevD.67.064009,
  title = {Spherically symmetric spacetimes in massive gravity},
  author = {Damour, Thibault and Kogan, Ian I. and Papazoglou, Antonios},
  journal = {Phys. Rev. D},
  volume = {67},
  issue = {6},
  pages = {064009},
  numpages = {25},
  year = {2003},
  month = {Mar},
  publisher = {American Physical Society},
  doi = {10.1103/PhysRevD.67.064009},
  url = {https://link.aps.org/doi/10.1103/PhysRevD.67.064009}
}

@article{PhysRevD.76.024003,
  title = {Numerical simulations of gravitational collapse in Einstein-aether theory},
  author = {Garfinkle, David and Eling, Christopher and Jacobson, Ted},
  journal = {Phys. Rev. D},
  volume = {76},
  issue = {2},
  pages = {024003},
  numpages = {9},
  year = {2007},
  month = {Jul},
  publisher = {American Physical Society},
  doi = {10.1103/PhysRevD.76.024003},
  url = {https://link.aps.org/doi/10.1103/PhysRevD.76.024003}
}

@article{Verbin:2020fzk,
    author = "Verbin, Y.",
    title = "{Magnetic and electric black holes in the vector-tensor Horndeski theory}",
    eprint = "2011.02515",
    archivePrefix = "arXiv",
    primaryClass = "gr-qc",
    doi = "10.1103/PhysRevD.106.024057",
    journal = "Phys. Rev. D",
    volume = "106",
    number = "2",
    pages = "024057",
    year = "2022"
}

@article{DamourEspositoFarese1993,
  author = {Damour, Thibault and Esposito-Far{\`e}se, Gilles},
  title = {Nonperturbative strong-field effects in tensor-scalar theories of gravitation},
  journal = {Phys. Rev. Lett.},
  volume = {70},
  pages = {2220--2223},
  year = {1993},
  doi = {10.1103/PhysRevLett.70.2220}
}

@article{LieblingPalenzuela2023,
  author = {Liebling, Steven L. and Palenzuela, Carlos},
  title = {Dynamical boson stars},
  journal = {Living Rev. Relativ.},
  volume = {26},
  pages = {1},
  year = {2023},
  doi = {10.1007/s41114-023-00043-4}
}

@article{Doneva2024Scalarization,
  author = {Doneva, Daniela D. and Ramazano{\u{g}}lu, Fethi M. and Silva, Hector O. and Sotiriou, Thomas P. and Yazadjiev, Stoytcho S.},
  title = {Spontaneous scalarization},
  journal = {Rev. Mod. Phys.},
  volume = {96},
  pages = {015004},
  year = {2024},
  doi = {10.1103/RevModPhys.96.015004}
}

@article{PhysRevD.40.2564,
  title = {Einstein gravity coupled to a massless scalar field in arbitrary spacetime dimensions},
  author = {Xanthopoulos, Basilis C. and Zannias, Thomas},
  journal = {Phys. Rev. D},
  volume = {40},
  issue = {8},
  pages = {2564--2567},
  numpages = {0},
  year = {1989},
  month = {Oct},
  publisher = {American Physical Society},
  doi = {10.1103/PhysRevD.40.2564},
  url = {https://link.aps.org/doi/10.1103/PhysRevD.40.2564}
}

@article{Song_2026,
   title={Bounds on the photon sphere radius for spherically symmetric black holes in n-dimensional Einstein gravity},
   volume={86},
   ISSN={1434-6052},
   url={http://dx.doi.org/10.1140/epjc/s10052-026-15623-1},
   DOI={10.1140/epjc/s10052-026-15623-1},
   number={4},
   journal={The European Physical Journal C},
   publisher={Springer Science and Business Media LLC},
   author={Song, Yong and Fu, Jiaqi and Cen, Yiting},
   year={2026},
   month=Apr }

\end{document}